\documentclass[epjCONF,columns]{svjour}
\usepackage{graphics}
\usepackage[varg]{txfonts} 
\usepackage[applemac]{inputenc}
\session-title{Hadron Collider Physics symposium 2011}
\begin{document}
\title{Search for WH associated production at D{\o} Tevatron}
\author{Florian Miconi\inst{1}\fnmsep\thanks{\email{Florian.Miconi@iphc.cnrs.fr}} }
\institute{Université de Strasbourg, IPHC, 23 rue du Loess 67037 Strasbourg, France \\
CNRS, UMR7178, 67037 Strasbourg, France}
\abstract{
The Higgs mechanism introduced in 1964 gives a satisfactory solution to a major problem of the standard model of elementary particles: the origin of the mass. It predicts the existence of the Higgs scalar boson, which has not been discovered experimentally yet.
The Tevatron, a hadron accelerator based at Fermi National Accelerator Laboratory near Chicago, has delivered data to its two multi-purpose detectors CDF and DZERO since 1983 up to september 2011. Leaving us about 11 $fb^{-1}$ of data per experiment to analyze.
Associated production of a Higgs boson and a vector gauge boson W or Z is the main search channel for a light standard Higgs boson (i.e. below 135 $GeV/c^2$). Using data collected by DZERO, we are looking for this production mode taking advantage of sophisticated techniques to improve the signal sensitivity such as b-jet identification and multivariate discriminants.
In the end, a statistical approach allows us to set an upper limit on the ratio between the observed (resp. expected) Higgs production cross section and its theoretical cross section. The latest result obtained in the WH channel using 8.5 $fb^{-1}$ at DZERO is 4.6 (resp. 3.5) for a 115 $GeV/c^2$ Higgs boson.
} 
\maketitle
\section{Introduction}
\label{intro}
\begin{figure}[!h]
\begin{center}
\resizebox{0.85\columnwidth}{!}{
 \includegraphics{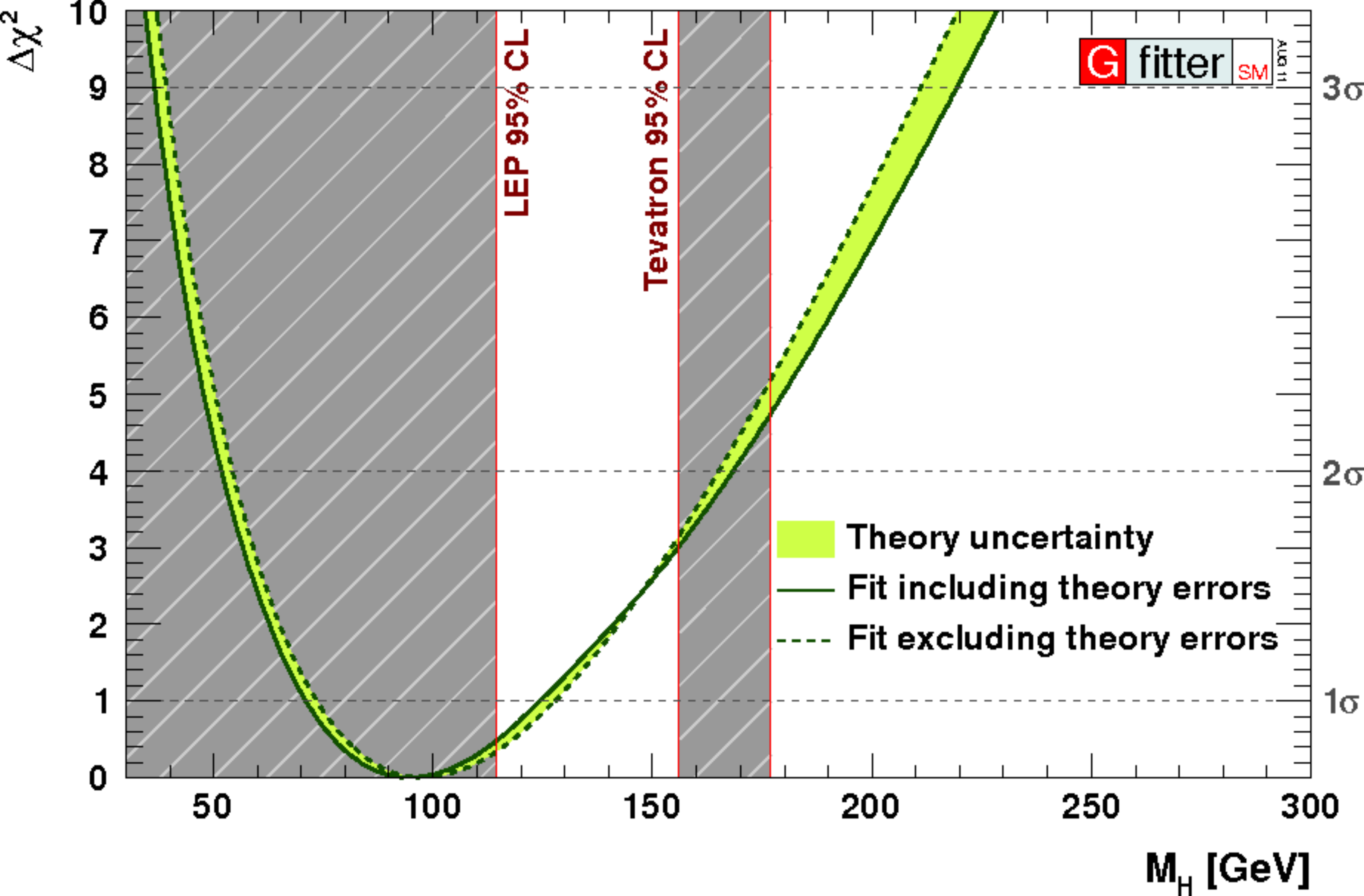}}
\caption{Direct and indirect experimental constraints on the Higgs boson mass.}
\label{fig:excl}       
\end{center}
\end{figure}

The mass of the Higgs boson is unpredicted by the Higgs mechanism of the standard model. Nevertheless, theoretical constraints exist but are very loose (considering triviality and vacuum stability). Combination of direct searches at colliders, and precision measurements of electroweak observables depending on the Higgs mass allows a more strict constraint over the mass range. We show on the figure \ref{fig:excl} the LEP exclusion up to 114.4 $GeV/c^2$ and the latest Tevatron exclusion ($156<m_H<177\ GeV/c^2$) along with the standard fit on the electroweak parameters measurements \cite{Baak:2011ze} . These results clearly show a preference for the low mass region.

Production decay modes are well established in the standard model as a function of the mass of the Higgs boson. Gluon fusion has the greatest cross section over all the mass region. But at low mass, the Higgs boson decays in two b quarks, making its signal impractical to extract within the overwhelming QCD background generated at a hadronic collider. Thus, the most sensitive channel are the WH and ZH associated production. The decay products of the W or the Z allowing to drastically reduce the hadronic background.

We present here the search for WH associated production at D{\o} Tevatron. D{\o} is one the two Tevatron’s multipurpose detector, designed to study high mass states and large transverse momentum phenomena, design that suits very well the search for a standard Higgs boson.
We will present the WH analysis strategy, starting in the section \ref{sec:selec} with its selection, we will then in the sections \ref{sec:bid} and \ref{sec:MVA} look at the techniques used to improve the signal efficiency by selecting the final state with b-jets identification algorithms and combining discriminant variables with multivariate analysis (MVA). We will also present in the section \ref{sec:3jets} how we partly recover radiation jets to improve the di-jet mass in the 3 jets channel using a radiation recovery method.

\section{The WH analysis selection}
\label{sec:selec}

The search is based on events with the following kinematical and topological characteristics :

\begin{itemize}
\item The reconstructed primary vertex must lay in the central region 60 cm around the center of the detector along the beam axis.
\item One central ($0\le \eta \le 2.5$) charged lepton (electron or muon) with a transverse momentum higher than 15 $GeV$ must be present.
\item We select events with a missing transverse energy above 15 $GeV$, signature of a neutrino coming from the decay of the W boson.
\item We consider only the events with 2 or 3 central ($0\le \eta \le 2.5$) jets with a transverse momentum higher than 20 $GeV$.
\item We also require a minimal value on a multivariate discriminant referred as MVA QCD aiming at reducing the QCD background.
\item Other leptons are vetoed in order to be orthogonal to the $ZH \rightarrow llbb$ analysis.
\end{itemize}

The analysis is then separated in two orthogonal channels with exactly two jets and three jets. These channels are further divided in orthogonal sub-channels with exactly one b-tagged jet and 2 b-tagged jets \cite{WH}.

\section{b-jet identification}
\label{sec:bid}

\begin{figure}[!h]
\begin{center}
\resizebox{0.70\columnwidth}{!}{
 \includegraphics{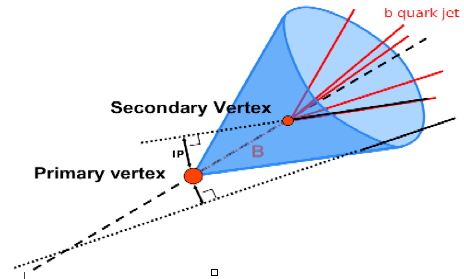} }
\caption{Diagram of a b-jet. IP corresponds to the track impact parameter}
\label{fig:bjet}       
\end{center}
\end{figure}

Being able to discriminate b-jets from lighter flavor jets is crucial for the low mass Higgs search. This tool relies on the characteristics of b quarks and B mesons. Indeed b-quark is the heaviest quark that lead to hadronisation before its decay and the resulting B-hadron has a very long lifetime (about 1.5 ps). These illustrated by figure \ref{fig:bjet} attributes lead to :

\begin{itemize}
\item A secondary vertex, about 3 mm apart from the primary vertex.
\item High impact parameter tracks produced by the B-hadron decay.
\item A higher secondary vertex mass and a distinctive jet angular opening.
\item The possible presence of a muon in the jet reconstruction cone.
\end{itemize}

\noindent Several algorithms have been developed at D\o\ to take advantage of these features \cite{btag}.

\begin{itemize}
\item SVT (Secondary Vertex Tagger) : based on the reconstruction of a secondary vertex using a Kalman filter.
\item JLIP (Jet LIfetime Probability tagger) : takes advantage of the impact parameter significance of tracks associated to a jet, to construct the probability that this jet is coming from the primary vertex and therefor more likely to be a light jet. Heavy flavor jets are distributed at low values of this variable.
\item CSIP (Counting Signed Impact Parameter tagger) : counts the number of tracks with high impact parameter.
\end{itemize}

\begin{figure}[!h]
\begin{center}
\resizebox{0.85\columnwidth}{!}{
 \includegraphics{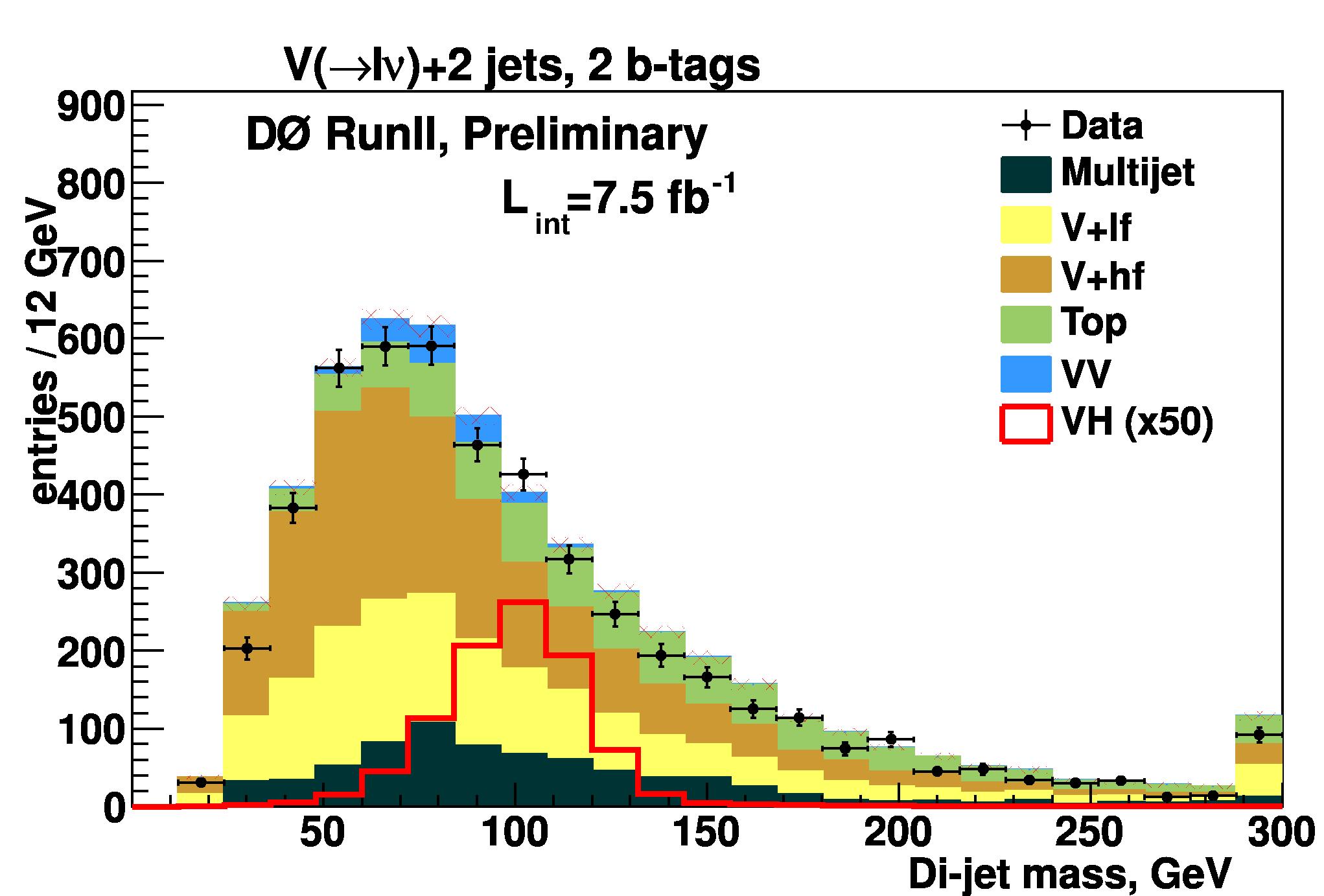} }
\caption{Higgs invariant mass for the 2 jets channel with 2 b-tagged jets}
\label{fig:mh}       
\end{center}
\end{figure}

Each of these algorithms show good performances in b-jets identification. As it has been observed that information given by each algorithm is not 100\% correlated since they don't exploit exactly the same b-jet features, it has been decided all these variables into a more powerful one using multivariate analysis. Contrarily to sequential cuts, a multivariate analysis allows to take into account the correlations between input variables.
On figure \ref{fig:mh} is shown the Higgs invariant mass reconstructed in the 2 jets channel with 2 jets b-tagged using the b-tagging multivariate analysis.

\section{Three jets channel special treatment: radiation recovery}
\label{sec:3jets}

\begin{figure}[!h]
\label{fig:mhvsrr}
\begin{center}
\resizebox{0.85\columnwidth}{!}{
 \includegraphics{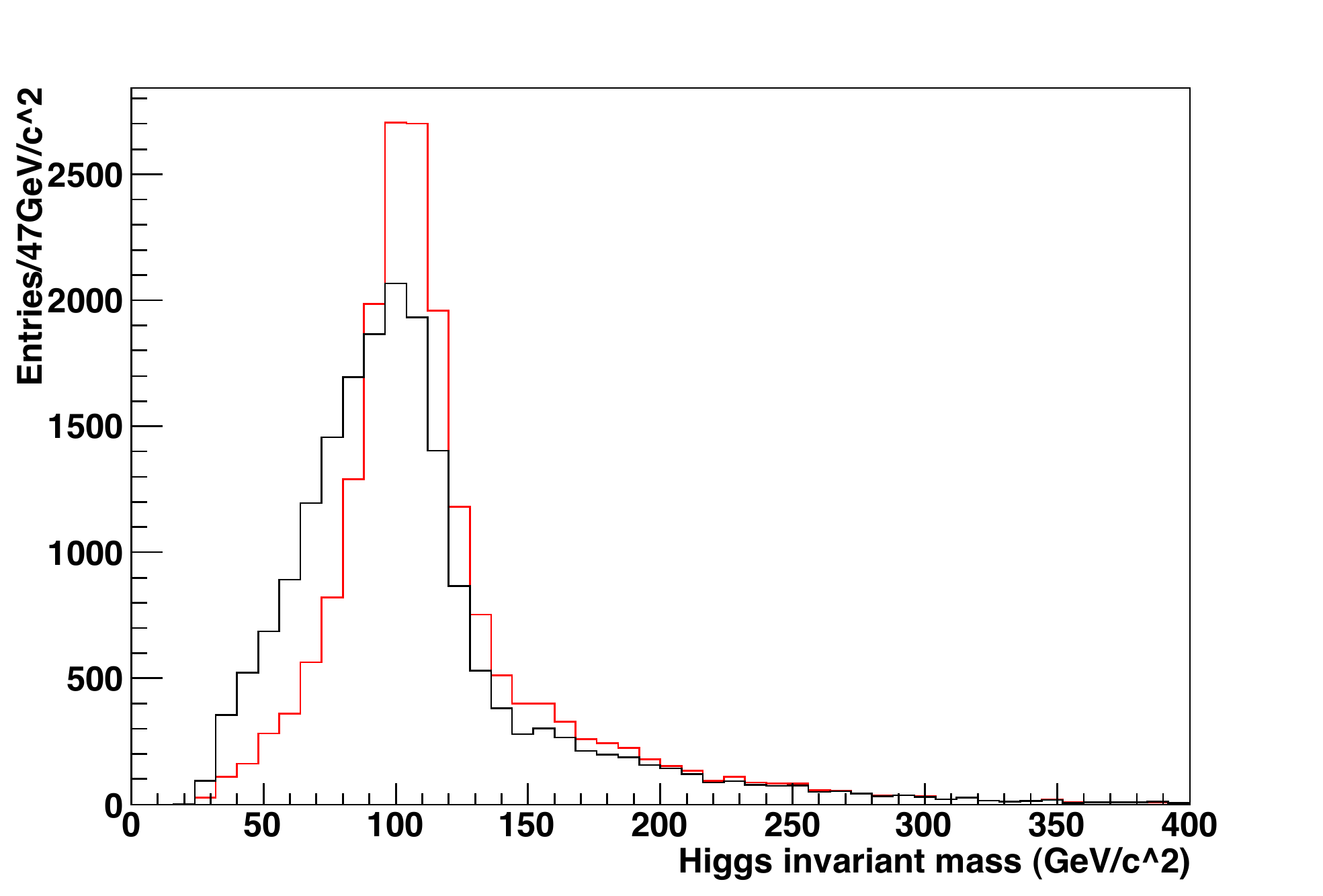} }
\caption{Higgs invariant mass with (red or light grey) and without (black) the radiation recovery method.}
\label{fig:1}       
\end{center}
\end{figure}

Higgs invariant mass is the most powerful single variable to search for a Higgs signal. In the 3 jets channel, its reconstruction is less straightforward than in the 2 jets channel. We want to take into account radiation coming from a b-jet (FSR) to reconstruct the Higgs invariant mass but we have to be careful not to use an initial state radiation (ISR) jet to do so. Figure \ref{fig:mhvsrr} shows Monte Carlo simulations of a 115 GeV Higgs, about 25\% improvement is achieved on the dijet-mass resolution if we recover every FSR jets. But in the detector, there's no simple way to discriminate a FSR jet from an ISR jet. We show that the minimal angle between the third and first or second jet (referred to {\it alphamin}) is efficient to discriminate these two cases.
In the end, the problem in a three jets selection is to find the right jet combination to reconstruct the Higgs invariant mass. The FSR recovery method consists in two steps :

\begin{itemize}
\item First, jets are ordered by decreasing b-jet identification variable value. This way, the two leading jets are the most likely to be the b-jets coming from the Higgs.
\item Knowing that the remaining jet is very likely to be the FSR or ISR jet, we add it to the invariant mass calculation if the value of {\it alphamin} is below an optimum cut that we found to be 1.
\end{itemize}

The method is currently implemented in most of the low mass analyses at D{\o} and shows up to 20\% improvement on the limit obtained in the 3 jets channel.

\section{Optimization using a multivariate analysis}
\label{sec:MVA}

\begin{figure}[!h]
\begin{center}
\resizebox{0.85\columnwidth}{!}{
 \includegraphics{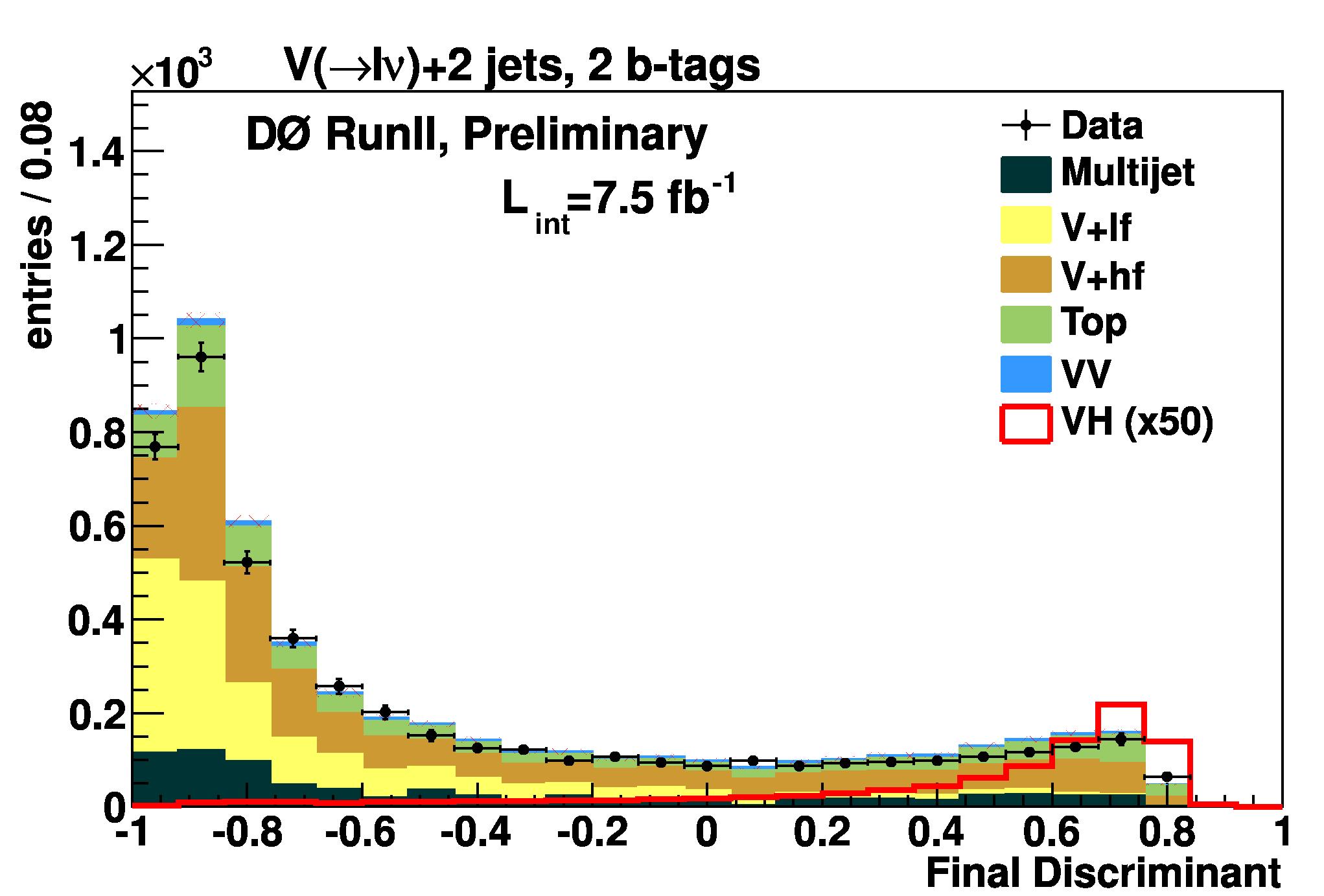} }
\caption{Final discriminant variable used in the 2 jets channel with 2 b-tagged jets}
\label{fig:mvafinal}       
\end{center}
\end{figure}

The same way multivariate analysis helped to achieve better performances of the b-tagging procedure, it can be very helpful to optimize the sensitivity to a Higgs signal by taking advantage of the discriminating power of less important variables and not only the invariant mass. The method chosen for the WH analysis is the Boosted Decision Tree (BDT) developed with the TMVA package \cite{TMVA}. A decision tree splits each event in order to optimize the signal/background separation. The resulting nodes (leaves) continue to be split until we reach the required signal/background purity. To select input variables we ensure that each is well-modeled and show good discriminating power between signal and background. A dedicated MVA is trained for each Higgs mass hypothesis covering the low mass region from 100 GeV to 150 GeV in 5 GeV steps.
Figure \ref{fig:mvafinal} shows the resulting output of the BDT in the 2 jets channel with 2 b-tagged jets for a 115 GeV Higgs, with the background events distributed towards 0 and well separated from the signal events.

\section{Extraction Limit}
\label{sec:limit}

\begin{figure}[!h]
\begin{center}
\resizebox{0.85\columnwidth}{!}{
 \includegraphics{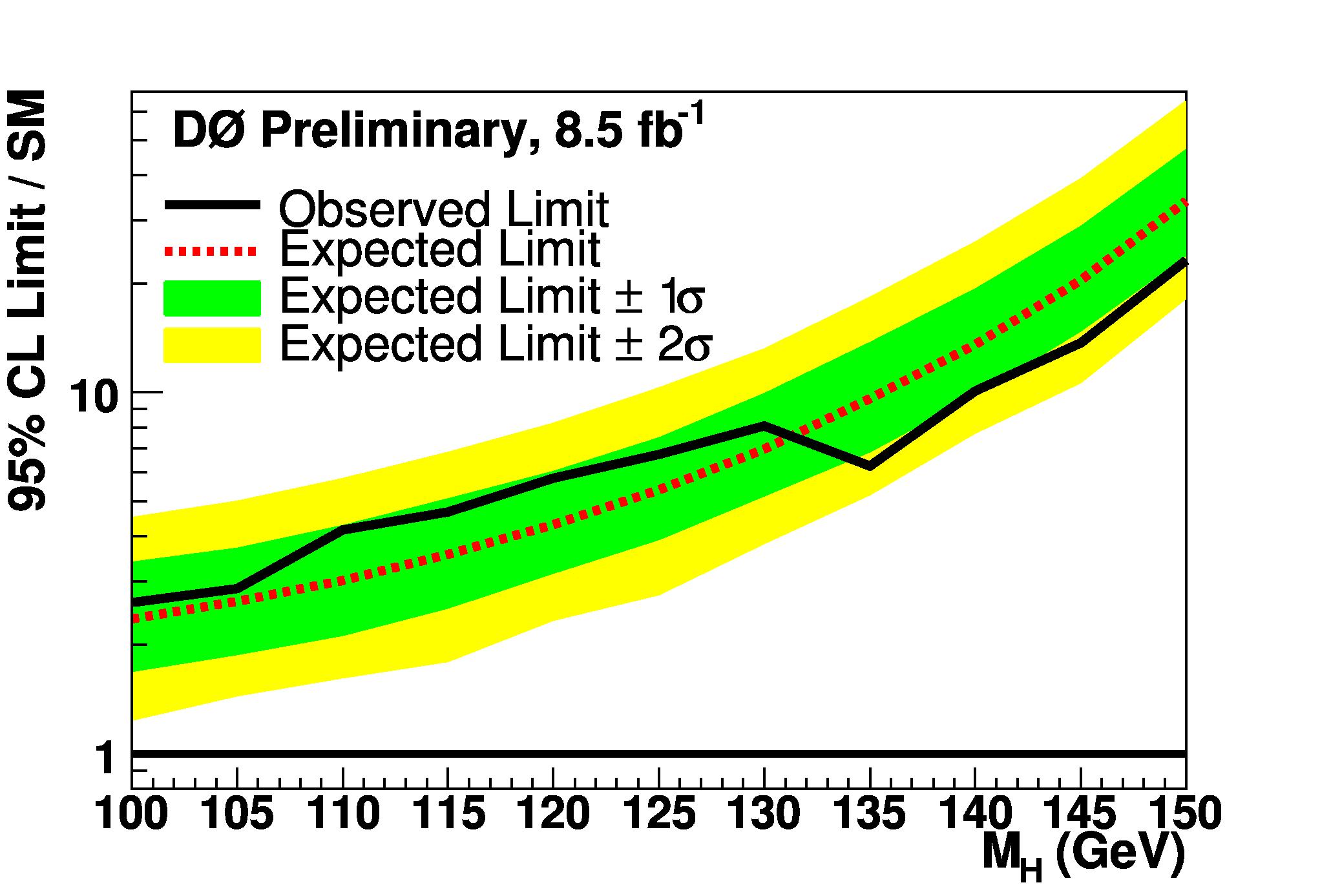} }
\caption{Limit on the Higgs production based on the WH analysis presented in this report.}
\label{fig:limitWH}       
\end{center}
\end{figure}

\begin{figure}[!h]
\begin{center}
 \resizebox{0.86\columnwidth}{!}{
 \includegraphics{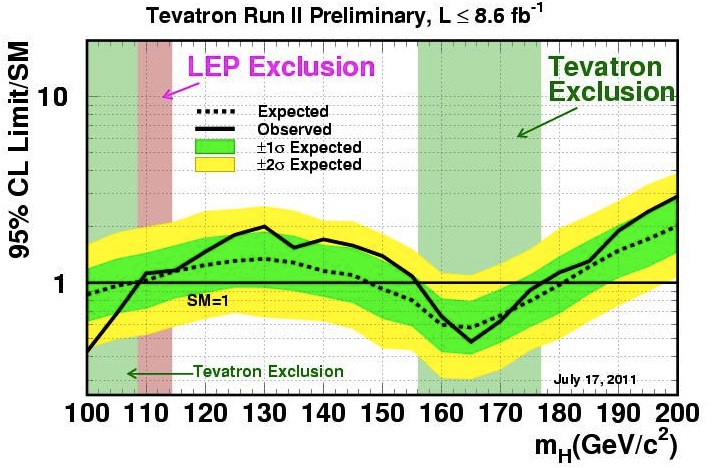} }
\caption{Limit on the Higgs production based on the combination of CDF and D{\o} results.}
\label{fig:limitTev}       
\end{center}
\end{figure}

The method used is based on the confidence levels method (CLs), that has already been used by the LEP collaborations. The algorithm developed in D{\o} is called Collie (Confidence Level Limit Evaluator), it is based on a semi-freq\-uentist statistical approach.
The method fully handles systematics uncertainties for a single result as well as for a combination of multiple channel and experiment results \cite{collie}. This method allows to set an upper limit on the ratio between the observed (resp. expected) Higgs production cross section and its cross section predicted by the standard model as a function of the Higgs mass. The latest result obtained in the WH channel using 8.5 fb-1 at Dø is shown on figure \ref{fig:limitWH}. Results combining all Tevatron analyses are very promising since the observed limit is below twice the standard model over all the low mass range as can be seen on the figure \ref{fig:limitTev} \cite{Aaltonen:2011gs}.

\end{document}